\documentclass[aps,prb,superscriptaddress,floatfix,twocolumn,showpacs]{revtex4-1}

\usepackage{verbatim}

\usepackage{amsmath}
\usepackage{amsfonts}
\usepackage{amssymb}
\usepackage{graphicx}
\usepackage{bm}
\usepackage{color}
\usepackage[hypertex]{hyperref}

\newcommand{\ket}[1]{|{#1}\rangle}
\newcommand{\bra}[1]{\langle{#1}|}

\newcommand{\HH}{\mathcal{H}}

\usepackage{graphicx}
\usepackage{bm}

\def\stacksymbols #1#2#3#4{\def\theguybelow{#2}
    \def\verticalposition{\lower#3pt}
    \def\spacingwithinsymbol{\baselineskip0pt\lineskip#4pt}
    \mathrel{\mathpalette\intermediary#1}}
\def\intermediary#1#2{\verticalposition\vbox{\spacingwithinsymbol
      \everycr={}\tabskip0pt
      \halign{$\mathsurround0pt#1\hfil##\hfil$\crcr#2\crcr
               \theguybelow\crcr}}}

\begin{document}
\title{
Unveiling hidden topological phases of a one-dimensional Hadamard quantum walk
}

\author{Hideaki Obuse}
\affiliation{
Department of Applied Physics, Hokkaido University, Sapporo 060-8628,
Japan}
\author{J{\'a}nos K. Asb{\'o}th}
\affiliation{
Institute for Solid State Physics and Optics,
Wigner Research Centre, Hungarian Academy of Sciences,
H-1525 Budapest P.O. Box 49, Hungary}
\author{Yuki Nishimura}
\affiliation{Department of Physics, Kyoto University, Kyoto 606-8502, Japan}
\author{Norio Kawakami}
\affiliation{Department of Physics, Kyoto University, Kyoto 606-8502, Japan}

\date{July 3, 2015}

\begin{abstract}
Quantum walks, whose dynamics is prescribed by alternating unitary
coin and shift operators, possess topological phases akin to those of
Floquet topological insulators, driven by a time-periodic
field. While there is ample theoretical work on topological phases of
quantum walks where the coin operators are spin rotations, in
experiments a different coin, the Hadamard operator is often used
instead. This was the case in a recent photonic quantum walk
experiment, where protected edge states were observed between two
bulks whose topological invariants, as calculated by the standard
theory, were the same. This hints at a \emph{hidden topological
  invariant} in the Hadamard quantum walk. We establish a relation
between the Hadamard and the spin rotation operator, which allows us
to apply the recently developed theory of topological phases of
quantum walks to the one-dimensional Hadamard quantum walk. The
topological invariants we derive account for the edge state observed
in the experiment, we thus reveal the hidden topological invariant of
the one-dimensional Hadamard quantum walk.
\end{abstract}


\maketitle

\section{Introduction}
Topological insulators have attracted much attention from various
branches of physics, due to their unique surface states predicted by
topological invariants\cite{rmp_kane,rmp_zhang}.  Although many
materials, such as HgTe, Bi$_2$Se$_3$, etc., have been identified as
being topological insulators, the necessary requirements on the intrinsic
parameters, e.g., the spin-orbit interactions and internal magnetic
fields, are hard to meet. One proposed way to overcome this
difficulty, is to use Floquet topological insulators, i.e., to employ
a periodic drive to bring a material from a topologically trivial
phase to the topologically nontrivial
one\cite{kitagawa_periodic,floquet_topological,
  dora_review,rudner_driven}. By altering the drive sequence, not only
the parameters, but even the relevant symmetries of the system
\cite{schnyder_original,kitaev2009advances,schnyder_tenfold} can be
tuned, offering a versatile route to topological insulators.

Since the Floquet topological insulator is defined by a unitary time
evolution operator, the spectral properties of its effective
Hamiltonian are characterized by the quasienergy, which has a $2\pi$
periodicity, in natural units where the time is measured in units of
the drive period and $\hbar = 1$.  As a consequence, in the presence
of chiral or particle-hole symmetries, surface states at $\pi$ as well
as 0 quasienergy should be considered. The number of $\pi$ quasienergy
states is a novel topologically protected quantity, and is a
corresponding novel bulk topological number. Accordingly,
1-dimensional chiral (or particle-hole) symmetric Floquet topological
insulators are characterized by two topological invariants, ${\mathbb
  Z}\times {\mathbb Z}$ (or ${\mathbb Z}_2 \times {\mathbb Z}_2$).
Thus, the Floquet topological insulator provides richer physics
compared with the time-independent topological insulator.

Floquet topological insulators have already been realized in various
experiments: by fabricating coupled helical waveguides  for laser pulses
in fused silica\cite{rechtsman_photonic_2013}, by irradiating the surface of a
topological insulator with circularly polarized
light\cite{wang2013observation}, and by ``shaking'', i.e.,
periodically modulating an optical lattice with trapped cold atoms to
realize the Haldane model\cite{jotzu2014experimental} and Hofstadter
model\cite{aidelsburger2013realization}.
However, none of these experiments has yet identified the unique
$\pi$ quasienergy states so far.

There is a promising way to realize Floquet topological insulators
using discrete-time quantum walks\cite{kempe_2003,ambainis_2003}
(quantum walks for short).  The dynamics of a quantum walk is
implemented by combining two fundamental operators: coin and shift
operators, which change the internal degree of freedom and the
position of a walker, respectively.  Recently, such quantum walks have
been experimentally realized in various systems, such as cold
atoms\cite{meschede_science}, trapped
ions\cite{schmitz_ion,roos_ions}, optical fiber
loops\cite{gabris_prl,schreiber_science}, bulk
optics\cite{white_photon_prl}, and integrated photonic
circuits\cite{sciarrino_twoparticle_natphot}.  It has been
clarified\cite{kitagawa_exploring,kitagawa_introduction} that the
discrete-time quantum walk is an ideal platform to construct Floquet
topological insulators because of the high tunability of relevant
symmetries\cite{schnyder_original,kitaev2009advances,schnyder_tenfold},
and the parameters which are essential to establish non-trivial
topological phases.  Motivated by this work, studies of the
topological phase of quantum walks have been
started\cite{obuse_delocalization,asboth_prb,asboth_2013,
  scattering_walk2014,asboth_edge_delocalization,asboth_edge_transport},
and their connection with the entanglement in these walks has also
been investigated\cite{chandrashekar}. Remarkably, edge states
originating in the non-trivial topological phase at zero and $\pi$
quasienergy have been observed in an experiment on a
one-dimensional photonic quantum
walk\cite{kitagawa_observation}. However, the bulk topological
invariants predicting these edge states were not identified: edge
states were observed at an interface between two regions with the same
topological number.

In the present work, we identify the hidden topological invariants of
the Hadamard quantum walk realized in the experiment of
Ref.~\onlinecite{kitagawa_observation}.  We generalize the approach
used for chiral symmetric quantum walks\cite{kitagawa_exploring,
  obuse_delocalization, kitagawa_introduction, asboth_prb,
  asboth_2013}, and find a pair of integers, i.e., a ${\mathbb
  Z}\times {\mathbb Z}$ topological invariant. 

This paper is organized as follows. In Sec.~\ref{sec:definitions}, we
define the one-dimensional discrete-time Hadamard quantum walk and
show how it is related to more commonly investigated quantum walks.
In Sec.~\ref{sec:chiral_symmetry} we define a generalization of chiral
symmetry for the Hadamard quantum walk, and give the formulas for the
corresponding chiral symmetric timeframes. In
Sec.~\ref{sec:topological number} we calculate the topological
invariants of the simple and the split-step Hadamard walks, and
illustrate the consequences of these invariants, the topologically
protected bound states, by numerical examples. In
Sec.~\ref{sec:experiment}, we apply this formalism to the setup
realized in Ref.~\onlinecite{kitagawa_observation}, and demonstrate
that the bound states observed in the experiment are predicted by our
bulk topological invariants. Finally, Sec.~\ref{sec:discussion} is
devoted to discussions and conclusion.

\section{Hadamard quantum walks}
\label{sec:definitions}

In this paper we consider one-dimensional quantum walks where the
walker has two internal states, denoted by $|+\rangle := (1,0)^{T}$ and
$|-\rangle:=(0,1)^T$. The wave function of the walker reads 
\begin{align}
\ket{\Psi(t)} &= \sum_{x\in\mathbb{Z}} \sum_{s = \pm} 
\Psi_{x,s}(t) \ket{x} \otimes \ket{s},
\end{align}
where $x\in \mathbb{Z}$ is the discrete position and $t\in\mathbb{N}$
is the discrete time.  The time evolution is generated by the unitary
timestep operator as
\begin{align}
\ket{\Psi(t)} &= U^t \ket{\Psi(0)}, 
\label{eq:standard U}
\end{align}
where the timestep operator $U$ is composed of a sequence of coin
operators $C$ and shift operators $S$, to be defined below.

A coin operator $C$ acts on the internal state of the walker while
leaving the position $x$ unaffected, 
\begin{equation}
C_H[\theta(x)] :=
\sum_x 
|x\rangle \langle x| \otimes \HH[\theta(x)].  
\label{eq:coin}
\end{equation}
We take as coin operator 
the generalized Hadamard operator,
\begin{equation}
\HH[\theta(x)] :=
\begin{pmatrix}
+\cos\theta(x) & +\sin\theta(x) \\
+\sin\theta(x) & -\cos\theta(x)
\end{pmatrix},
\label{eq:Hadamard}
\end{equation}
with a parameter $\theta$ depending on the position $x$.
This can be expressed using the Pauli matrices
\begin{eqnarray*}
 \sigma_1 := 
 \begin{pmatrix}
0 & 1\\
1 & 0
 \end{pmatrix},
\sigma_2 := 
 \begin{pmatrix}
0 & -i\\
i & 0
 \end{pmatrix},
\sigma_3 := 
 \begin{pmatrix}
1 & 0\\
0 & -1
 \end{pmatrix},
\end{eqnarray*}
and the identity matrix $\sigma_0 := {\bf I}_2$, 
as 
\begin{eqnarray}
 \HH[\theta] = e^{-i\theta \sigma_2}\sigma_3 = \cos\theta \sigma_3
 + \sin \theta \sigma_1.
\label{eq:Hadamard-rotation}
\end{eqnarray}
Most of the previous theoretical work used as coin operator the
rotation $e^{-i\theta \sigma_2} = \cos\theta \sigma_0 -i \sin \theta
\sigma_2$, which has the same matrix elements as the Hadamard operator
up to the position of the minus sign. Although this seems like a small
difference, it can have far reaching consequences, as we will show
below.

A shift operator is complementary to the rotation operator in that it changes
the position of the walker in a way that depends on the value of the
internal degree of freedom. We will use two shift
operators, defined as  
\begin{align}
\label{eq:shift}
 S_\pm &:= \sum_x\big(
|x\pm1\rangle \langle x| \otimes |\pm\rangle \langle \pm|
+
|x\rangle \langle x| \otimes |\mp\rangle \langle \mp|
\big);
\end{align}

We consider two types of quantum walks, constructed from the Hadamard
coin and the shift operators above.  
The simple Hadamard quantum walk
is defined via its timestep operator as
\begin{equation}
 U_A(\theta(x)) = 
S_- S_+
 C_\text{H}(\theta(x)).
\label{eq:U_single}
\end{equation}
The split-step Hadamard walk has the timestep operator
\begin{equation}
 U_B[\theta_1(x), \theta_2(x)] = S_- C_\text{H}[\theta_2(x)] S_+
 C_\text{H}[\theta_1(x)].
\label{eq:U_split}
\end{equation}

For both the simple and the split-step quantum walk, the effective
Hamiltonian $H$ is a useful tool to understand their long-time
dynamics.  It is defined from the unitary timestep operator by
\begin{equation}
U= e^{-i H}. 
\label{eq:U H}
\end{equation}
Stationary states of a quantum walk are eigenstates of the timestep
operator $U$, 
\begin{equation}
 U \ket{\psi_\varepsilon} = \lambda_\varepsilon
 \ket{\psi_\varepsilon}, \quad \lambda_\varepsilon=e^{- i \varepsilon}.
\end{equation}
Here the \emph{quasienergy} $\varepsilon$, the eigenenergy of the
effective Hamiltonian, has $2 \pi$ periodicity: due to unitarity of
$U$, $\lambda_\varepsilon$ takes its values from the unit circle on
the complex plane.

\section{Chiral symmetry of Hadamard quantum walks}
\label{sec:chiral_symmetry}

For a one-dimensional Hamiltonian to possess topological phases, it
needs to have some symmetry that links positive and negative energy
states to each
other\cite{schnyder_original,kitaev2009advances,schnyder_tenfold}. We
suggest an extension of the concept of chiral symmetry, and show that
both Hadamard walks possess it. This will later allow us to describe
the bulk topology and protected edge states of the Hadamard walks.

\subsection{Chiral symmetry at nonzero energy}

As a starting point we introduce the \emph{chiral symmetry at
nonzero energy} for a Hamiltonian.  Consider a system of free fermions,
with grand canonical Hamiltonian
\begin{align}
\hat{H} &= \sum_{nm} \hat{c}^\dagger_n H_{nm}(\xi) \hat{c}_m - 
\mu \sum_n \hat{c}_n^\dagger \hat{c}_n,
\end{align}
where the matrix of the single-particle Hamiltonian $H$ is a
continuous function of some system parameters denoted by $\xi
\in \Xi$. This includes all parameters that are subject to
disorder. 
The requirement for chiral symmetry of the Hamiltonian reads
\begin{align}
\Gamma H(\xi) \Gamma &= -H(\xi),
\label{eq:chiral Hamiltonian}
\end{align}
with a unitary chiral symmetry operator $\Gamma =\Gamma^\dagger =
\Gamma^{-1}$, that acts in each unit cell independently, and is
independent of the disorder realization $\xi$. 

A {\it global, fixed} onsite potential $\phi \in \mathbb{R}$, that is not
subject to disorder,
\begin{align}
H_{nm}'(\xi) = H_{nm}(\xi) + \phi, 
\end{align}
obviously breaks chiral symmetry, since $\phi \in \mathbb{R}$ commutes
with any $\Gamma$ instead of anticommuting.  However, 
all it does is simply displace the energy of all states by
$\phi$. Thus, if $H$ hosts topologically protected bound states, so
will $H'$, the only difference is that
they will be at energy $\phi$ instead of energy $0$.

The same discussion applies to periodically driven systems. The
requirement of Eq.\ (\ref{eq:chiral Hamiltonian}), translated for the
time-evolution operator using Eq.~\eqref{eq:U H}, but allowing for a
constant shift of quasienergy, reads
\begin{equation}
 \Gamma U(\xi) \Gamma = e^{-2i\phi} U(\xi)^{-1}.
\label{eq:chiral symmetry}
\end{equation}
Importantly, not only $\Gamma$, but also $\phi \in \mathbb{R}$ is here
assumed not to be subject to disorder, i.e., independent of the
parameters $\xi$. If Eq.~\eqref{eq:chiral symmetry} holds, the
operator $e^{i\phi}U(\xi)$ has chiral symmetry (in the usual
sense\cite{asboth_2013,asboth2014chiral}), and it may host
topologically protected edge states at $\varepsilon=0$ or $\pi$.  If
this is the case, the original timestep operator $U(\xi)$ will have the
same topologically protected end states at quasienergy
$\varepsilon=\phi$, respectively, $\varepsilon=\pi+\phi$.  In the
following we will not write out the arguments $\xi$ representing the
effects of disorder explicitly. 

\subsection{Chiral symmetry of Hadamard quantum walks}

We follow the method developed by some of the authors of this
work\cite{asboth_2013} to describe chiral symmetry of quantum walks,
adapted to deal with chiral symmetry at finite quasienergy. A quantum
walk has chiral symmetry at finite quasienergy if its timestep
operator can be split into two parts, conjugated inverses of each
other:
\begin{align}
U &= e^{-i\phi} F \cdot \Gamma F^{-1} \Gamma, 
\label{eq:FG}
\end{align}
with $\phi\in\mathbb{R}$.  We can directly confirm chiral symmetry of
the above quantum walk by substituting $U$ and $U^{-1}$ in
Eq.\ (\ref{eq:FG}) into left and right hand sides in
Eq.\ (\ref{eq:chiral symmetry}), respectively, and using
$\Gamma^{-1}=\Gamma$.  In order to find such a decomposition, the
starting time of the period can be shifted, i.e., the walk can be
described in a different \emph{timeframe}. This process is detailed in
Ref.~\onlinecite{asboth_2013} for quantum walks where the coin
operator is a rotation, and where
\begin{equation}
 \Gamma :=  \sum_x |x\rangle \langle x | \otimes \sigma_1.
\label{eq:Gamma sigma_1}
\end{equation}

To reveal the chiral symmetry of the Hadamard walks, we start by
rewriting Eq.~\eqref{eq:Hadamard-rotation} as
\begin{align}
\HH[\theta(x)]=
e^{-i\theta(x) \sigma_2 }\cdot e^{-i\phi} \cdot e^{-i\chi \sigma_3}
\label{eq:def_chi_phi}
\end{align}
with $\chi =-\phi= \pi/2$. Importantly, $\chi$ and $\phi$ are {\it
  global, fixed} parameters. Now as for the factor $e^{-i\phi}$, it
commutes with all operators, and so it will only shift the quasienergy. 

To deal with the factor $e^{-i\chi \sigma_3}$ in
Eq.~\eqref{eq:def_chi_phi}, we first make a few observations.  It can
be broken up into two parts, as
\begin{align}
e^{-i\chi \sigma_3} &= e^{-i\chi/2 \cdot \sigma_3} \cdot e^{-i\chi/2
 \cdot \sigma_3}. 
\label{eq:break sigma_3}
\end{align}
Then we notice that $e^{-i\chi/2 \cdot \sigma_3}$ commutes with both 
$S_+$ and $S_-$, 
since 
\begin{align*}
 S_{\pm} &= \sum_x \Big(\, \ket{x\pm 1}\bra{x} \otimes
   \frac{\sigma_0 \pm \sigma_3}{2} + 
\ket{x}\bra{x} \otimes \frac{\sigma_0 \mp \sigma_3}{2}\,
\Big).
\end{align*}
Finally, we point out the relation
\begin{align}
\sigma_1 \cdot e^{-i\chi/2 \cdot \sigma_3 } \cdot\sigma_1 &= 
e^{i\chi/2 \cdot \sigma_3 } = (e^{-i\chi/2 \cdot \sigma_3 })^{-1}, 
\label{eq:chiral sigma_3}
\end{align}
which will be useful to show chiral symmetry. 

Using the results of the previous paragraph, we rewrite the timestep
operators of the Hadamard walks, Eqs.~\eqref{eq:U_single} and
\eqref{eq:U_split} in a timeframe where the chiral symmetry at nonzero
quasienergy is explicit. For the simple Hadamard walk this reads
\begin{subequations}
\begin{align}
U_A &=  e^{-i\phi} F_A \cdot \Gamma F_A^{-1} \Gamma;
\label{eq:UA_single_chiral}\\
\label{eq:F_single_chiral}
F_A &= 
e^{-i\theta(x)/2 \cdot \sigma_2} \cdot
e^{-i\chi/2 \cdot \sigma_3 } 
S_-
;\\
\Gamma F_A^{-1} \Gamma &= 
S_+ \cdot
e^{-i\chi/2 \cdot \sigma_3 }  \cdot
e^{-i\theta(x)/2 \cdot \sigma_2 }
,
\end{align}
\label{eq:U_single_chiral}
\end{subequations}
while for the split-step Hadamard walk we find 
\begin{subequations}
\label{eq:U_split_chiral}
\begin{align}
U_B &=  e^{-i 2\phi} F_B \cdot \Gamma F_B^{-1} \Gamma;
\label{eq:UB_split_chiral} \\
\label{eq:F_split_chiral}
F_B &= 
e^{-i\theta_1(x)/2 \cdot \sigma_2 } 
S_- \cdot
e^{-i\chi \sigma_3} \cdot
e^{-i\theta_2(x)/2 \cdot \sigma_2 },
\\
\Gamma F_B^{-1} \Gamma &= 
e^{-i\theta_2(x)/2 \cdot \sigma_2 } \cdot
e^{-i\chi \sigma_3} 
S_+ \cdot
e^{-i\theta_1(x)/2 \cdot \sigma_2 },
\end{align}
\end{subequations}
with, in both cases, $\chi= -\phi = \pi/2$.  We remark that chiral
symmetry of the simple and split-step Hadamard walks is preserved even
when the parameter $\theta(x)$ of the coin operator in
Eq.\ (\ref{eq:Hadamard}) depends on the position $x$ in a disordered
way.

\section{Topological phases of Hadamard quantum walks}
\label{sec:topological number}

Having established chiral symmetry for the Hadamard quantum walks, we
now determine the bulk topological invariants controlling the number
of edge states in these walks\cite{asboth_2013, scattering_walk2014,
  asboth2014chiral}.  We will follow the procedure developed in
Ref.\ \onlinecite{asboth2014chiral}, which expresses the topological
invariants as winding numbers of parts of the operator $F$ from
Eq.~\eqref{eq:FG} between eigenspaces of the chiral symmetry operator
$\Gamma$.  In Appendix \ref{sec:appendix}, an alternative procedure
developed in Ref.\ \onlinecite{asboth_2013} is presented. 

To briefly summarize, Ref.\ \onlinecite{asboth2014chiral} states that
\begin{subequations}
\label{eq:nu0pi_def}
\begin{align}
\nu_0 &= \frac{1}{2\pi i} \int_{-\pi}^\pi dk \frac{d}{dk} \text{ln } \text{det} F_{+-}(k);\\
\nu_\pi &= \frac{1}{2\pi i} \int_{-\pi}^\pi dk \frac{d}{dk} \text{ln } \text{det} F_{--}(k),
\end{align}
\end{subequations}
where $F_{+-}(k)$ is the part of $F$ in the quasimomentum space
representation that maps from the subspace of the
Hilbert space where $\Gamma=-1$  (i.e., the eigenspace of $\Gamma$
belonging to eigenvalue $-1$) to the $\Gamma=+1$ subspace, while $F_{--}$ is
the part of $F$ that acts in the $\Gamma=-1$ subspace.

To adapt the results of Ref.\ \onlinecite{asboth2014chiral} to the
Hadamard quantum walks, we need to take two things into account.
First, Eqs.~\eqref{eq:F_single_chiral} and \eqref{eq:F_split_chiral}
give the matrix of $F$ in a basis where the chiral symmetry operator
is not diagonal, that is $\Gamma=\sigma_1$. In such a basis, i.e.,
whenever
\begin{align}
\label{eq:gamma_sigmax_F_def}
\Gamma &= 
\begin{pmatrix}
0 & 1  \\
1 & 0  
\end{pmatrix}; \quad
F(k) = 
\begin{pmatrix}
a(k) & b(k) \\
c(k) & d(k)
\end{pmatrix},
\end{align}
for the parts of $F$ necessary for the topological invariants we have
\begin{subequations}
\label{eq:F_components}
\begin{align}
2 F_{+-} &= 
\begin{pmatrix}1 &\,\,\, 1\end{pmatrix}
  \begin{pmatrix} a & b \\ c & d \end{pmatrix} 
  \begin{pmatrix}\,\,\,\, 1 \\ -1 \end{pmatrix} = 
\left(a - b + c - d \right),\\
2 F_{--} &= 
\begin{pmatrix}1 & \!-1\end{pmatrix}
  \begin{pmatrix} a & b \\ c & d \end{pmatrix} 
  \begin{pmatrix}\,\,\,\, 1 \\ -1 \end{pmatrix} = 
\left(a - b - c + d \right).
\end{align}
\end{subequations}
Second, the Hadamard walks have chiral symmetry at finite
quasienergy $\phi$. Instead of topological invariants $\nu_0$ and
$\nu_\pi$, we thus have invariants $\nu_{\phi}$ and $\nu_{\pi+\phi}$,
which read 
\begin{subequations}
\label{eq:nu_phi_pi_def}
\begin{align}
\nu_{\phi} &= \frac{1}{2\pi i} \int_{-\pi}^\pi dk 
\frac{d}{dk} \text{ln } \left( a(k) - b(k) + c(k) - d(k) \right);
\\
\nu_{\pi+\phi} &= \frac{1}{2\pi i} \int_{-\pi}^\pi dk 
\frac{d}{dk} \text{ln } \left( a(k) - b(k) - c(k) + d(k) \right).
\end{align}
\end{subequations}
We obtained this by substituting Eqs.~\eqref{eq:F_components} into
Eqs.~\eqref{eq:nu0pi_def}, and omitting a factor of
$1/2$, which does not change the winding number.

\subsection{Simple Hadamard walk}

\begin{figure*}[t]
\includegraphics[width=14cm]{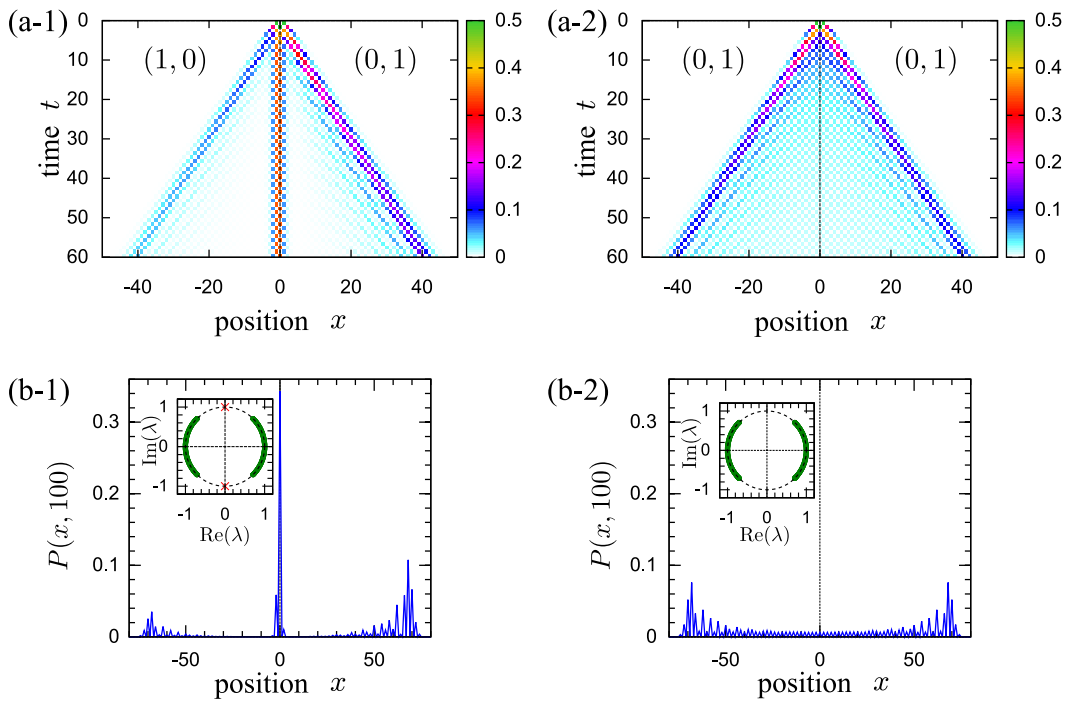}
\caption{(Color online) Examples for bound states as signatures of a
  topological phase boundary in single step Hadamard walks. Top:
  Contour maps of the probability distribution $P(x,t)$ in $x-t$ plane
  for the quantum walks of (a-1) $U_A({\bm
    \theta}_\alpha)$, Eq.~\eqref{eq:example_singlestep_alpha}, and
  (a-2) $U_A({\bm \theta}_\beta)$,
  Eq.~\eqref{eq:example_singlestep_beta}.  The topological numbers
  $(\nu_{-\pi/2},\nu_{+\pi/2})$ for the negative and positive $x$
  regions are shown in the figures. The position dependence of the
  probability distribution $P(x,t)$ at $t=100$ is shown in (b-1) and
  (b-2) for the single-step Hadamard walk $U_A({\bm \theta}_\alpha)$
  and $U_A({\bm \theta}_\beta)$, respectively.  The insets in (b-1)
  and (b-2) show eigenvalues $\lambda=e^{-i\varepsilon}$ of the
  corresponding time-evolution operator $U_A({\bm
    \theta}_{\alpha,\beta})$ of the main figure.  The eigenvalues
  corresponding to edge states are distinguished by red crosses from
  those of bulk states (green thick arcs).  }
\label{fig:Probability_Usingle}
\end{figure*}

We first consider the simple Hadamard quantum walk, as defined in
Eq.~\eqref{eq:U_single}.  
We take a translation invariant bulk,
with $\theta(x)=\theta$,  
in a chiral timeframe, as defined by Eq.~\eqref{eq:U_single_chiral}. 
The operator $F_A$ at quasimomentum $k$ reads 
\begin{align}
F_A(k) &= e^{-i\chi/2} 
e^{-i\theta/2 \cdot \sigma_2}
\begin{pmatrix}
1&0\\ 0& e^{i (k+\chi)} 
\end{pmatrix}.
\label{eq:F_single}
\end{align}
The parameter $\chi=\pi/2$ shows up in two roles here. First, it works as a
global phase, $k$-independent factor. This cannot change the winding
number. The second role is as a displacement of $k$ by $\chi$. This again does
not change the winding number which is calculated by integrating over
the whole $k$ space. Thereby, we are free to set $\chi=0$, when we
substitute Eq.\ (\ref{eq:F_single}) into Eqs.~\eqref{eq:F_components} and 
\eqref{eq:nu_phi_pi_def}. We obtain
\begin{subequations}
\begin{align}
\nu_{-\pi/2} &= \frac{1}{2\pi i} \int_{-\pi}^{\pi} dk
\frac{d}{dk} \text{ln } 
\left(s_{\theta+}\, - e^{ik} s_{\theta-}  \right);\\
\nu_{+\pi/2} &= \frac{1}{2\pi i} \int_{-\pi}^{\pi} dk 
\frac{d}{dk} \text{ln } 
\left( s_{\theta-}\, + e^{ik} s_{\theta+}  \right),
\end{align}
\end{subequations}
using the shorthand
\begin{align}
\label{eq:s_pm_def}
s_{\theta\pm} &= \cos\frac{\theta}{2} \pm \sin\frac{\theta}{2}= 
\sqrt{2} \sin \frac{\theta\pm\pi/2}{2}.
\end{align}
The invariant at quasienergy $\pm\pi/2$ is the winding number of a
loop on the complex plane, centered at $s_{\theta \pm}$ with radius
$s_{\theta \mp}$. In the case  in which the radius is larger than the distance of
the center from the origin, the loop encircles the origin, and we have
a winding number of $+1$. In the opposite case, the winding number is
0. Therefore, to calculate the values of the winding numbers, we need
to consider
\begin{align}
s_{\theta+}^2 - s_{\theta-}^2 &= 2\sin \theta. 
\end{align}
For the winding numbers, the above considerations give
\begin{align}
(\nu_{-\pi/2}, \nu_{+\pi/2}) = \left\{
\begin{array}{rl}
(0, 1) & \quad \text{if} \quad 0<\theta<\pi, \\
(1, 0) & \quad \text{if} \quad -\pi<\theta<0.
\label{eq:nu for the single step Hadamard}
\end{array}
\right.
\end{align}
The topological invariants are not defined when $\theta=0$ or
$\theta=\pi$. In these cases, the time-evolution operator reads
$U_A (\theta = \pi/2 \pm \pi/2) = 
\pm i e^{-i (k+\pi/2) \sigma_3}, 
$
with a quasienergy spectrum that has no gaps. 

\subsubsection*{Numerical examples}

We illustrate the above results on the topology of the simple Hadamard
walk, by showing examples where edge states appear near a boundary
between two bulks with different topological invariants. To this end,
we define two sets of $\theta(x)$:
\begin{align}
{\bm \theta}_\alpha&: \{ \theta_- = -\pi/4,\ \theta_+ = +\pi/4\},
\label{eq:example_singlestep_alpha}\\
{\bm \theta}_\beta&: \{\theta_- = +3\pi/4,\  \theta_+ = +\pi/4\},
\label{eq:example_singlestep_beta}
\end{align}
where 
\begin{equation*}
\theta_- := \theta(x\le -1),  \quad\theta_+ := \theta(x\ge0).
\label{eq:theta_- theta_+}
\end{equation*}
According to Eq.\ (\ref{eq:nu for the single step Hadamard}), we
expect that topologically protected edge states at quasienergies
$\varepsilon=\pm \pi/2$ appear near $x \approx 0$ for $U_A({\bm
  \theta}_\alpha)$, because the two bulk regions have different
topological numbers, while no edge state should appear for $U_A({\bm
  \theta}_\beta)$ because of the same topological numbers in both
regions.

We numerically simulate the time evolution $|\Psi(t)\rangle = U_A^t
|\Psi(t=0)\rangle$ up to $t=100$ and calculate the probability
distribution
\begin{equation}
P(x,t) := \sum_{s=+,-}|(\langle x|\otimes\langle s |)|\Psi(t)\rangle|^2.
\end{equation}
The initial state is set to
\begin{equation}
 |\Psi(t=0)\rangle := (|+\rangle + i |-\rangle)/\sqrt{2} \otimes |0\rangle.
\end{equation}
Figures \ref{fig:Probability_Usingle} (a-1) and (a-2) show the contour
maps of $P(x,t)$ in the $x-t$ plane up to the time step $t=60$ for
$U_A({\bm \theta}_\alpha)$, Eq.~\eqref{eq:example_singlestep_alpha},
and $U_A({\bm \theta}_\beta)$, Eq.~\eqref{eq:example_singlestep_beta}
respectively.  On the one hand, Fig.\ \ref{fig:Probability_Usingle} (a-1)
clearly shows that the high probability amplitudes stably remain near
the origin where the topological numbers change.  On the other hand,
Fig.\ \ref{fig:Probability_Usingle} (a-2) exhibits low probability
amplitudes near $x=0$ as expected.

The presence/absence of edge states is further highlighted in
Figs.\ \ref{fig:Probability_Usingle} (b-1) and (b-2) showing snapshots
of the probability distribution $P(x,t)$ at $t=100$ for $U_A({\bm
  \theta}_\alpha)$ and $U_A({\bm \theta}_\beta)$, respectively.  We
also numerically compute the spectrum of the timestep operators of the
single-step Hadamard walks $U_A({\bm \theta}_\alpha)$ and $U_B({\bm
  \theta}_\beta)$, and show the eigenvalues
$\lambda_\varepsilon=e^{-i\varepsilon}$ in the insets.  Here we
consider the finite position space from $-L$ to $L-1$ with $L=100$ and
impose the periodic boundary conditions to $-L$ and $L-1$.  Thereby,
we have two boundaries at $x=0$ and $-L$, where $\theta(x)$ is varied.
If one of these boundaries hosts an edge state at quasienergy
$\varepsilon$, so must the other boundary: this gives an extra double
degeneracy of the edge states. 
Although this degeneracy is lifted, because wavefunctions of the edge
states at the same quasienergy but opposite edges overlap due to their
exponential tails, this is the
correction that is exponentially small in the system size, in our
case, below the numerical accuracy $\sim 10^{-16}$.  Consistent with
our theoretical prediction, the eigenvalues corresponding to edge
states (red crosses) appear at $\varepsilon=\pm \pi/2$ for $U_A({\bm
  \theta}_\alpha)$, while they are not there for $U_A({\bm
  \theta}_\beta)$.  This illustrates the validity of Eq.~(\ref{eq:nu
  for the single step Hadamard}).

\begin{figure}[t]
\includegraphics[width=7cm]{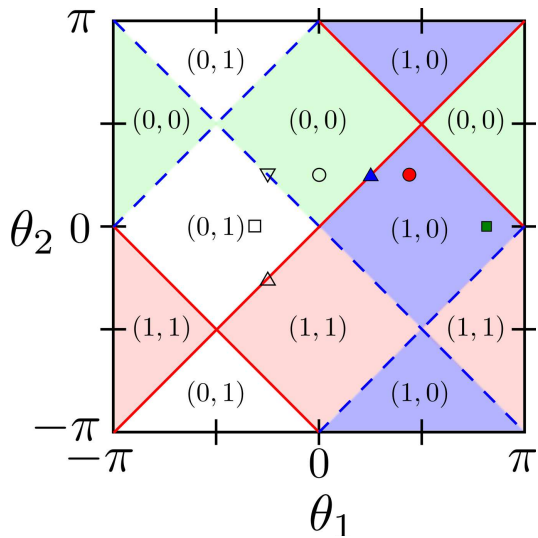}
\caption{(Color online) The phase diagram of the split-step quantum
  walk $U_B(\theta_1,\theta_2)$ defined in Eq.\ (\ref{eq:U_split}).
  Gapped phases are indexed by topological numbers $(\nu_0, \nu_\pi)$,
  Eqs.~\eqref{eqs:nu_splitstep}. The red solid and blue dashed lines
  indicate closing of quasienergy gaps around $\varepsilon=0$ and
  $\pi$, respectively.  The symbols in the phase diagram indicate the
  parameters of the numerical examples we consider, i.e., upper
  and lower triangles for ${\bm \theta}_\gamma$, ${\bm \theta}_\delta$
  in Eqs.~\eqref{eq:splitstep_example_delta} and
  \eqref{eq:splitstep_example_gamma}, circles for ${\bm \theta}_{e1}$
  in Eq.\ (\ref{eq:theta_exp1}), and rectangles for ${\bm
    \theta}_{e2}$ in Eq.\ (\ref{eq:theta_exp2}).
}
\label{fig:phase-diagram}
\end{figure}

\subsection{Split-step Hadamard walk}

\begin{figure*}[t]
\includegraphics[width=14cm]{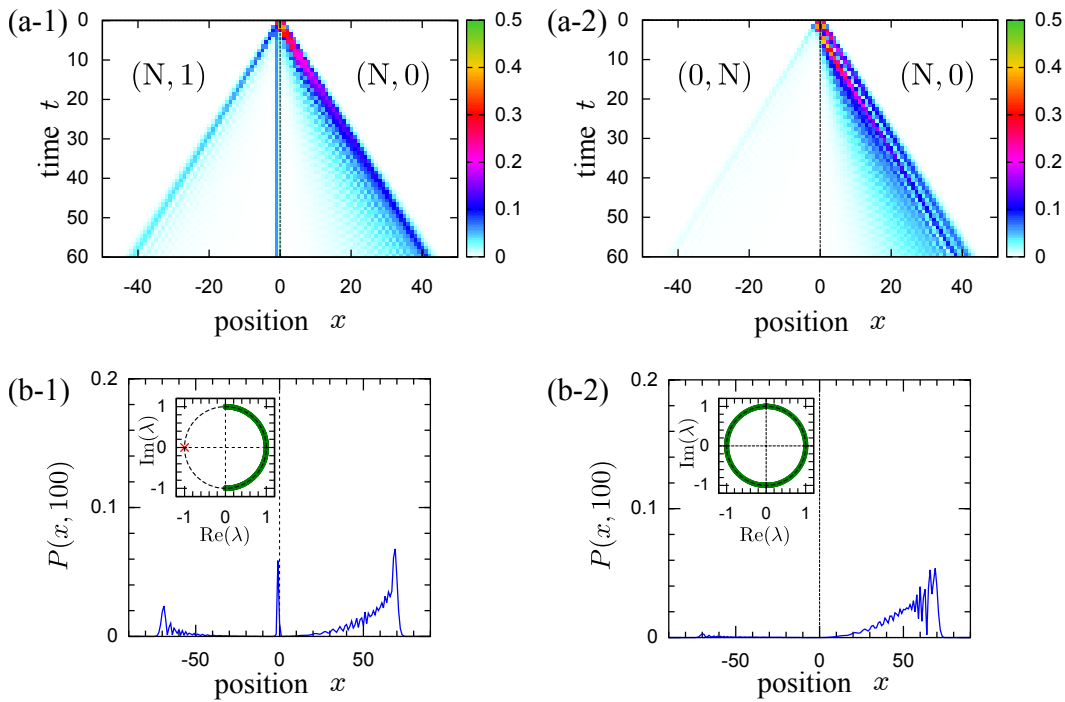}
\caption{(Color online) Examples of bound states as signatures of a
  topological phase transition in split-step Hadamard walks. Top:
  Contour maps of the probability distribution $P(x,t)$ in the $x-t$
  plane for the quantum walk (a-1) $U_B({\bm \theta}_\gamma)$,
  Eq.~\eqref{eq:splitstep_example_gamma} and (a-2) $U_B({\bm
    \theta}_\delta)$, Eq.~\eqref{eq:splitstep_example_delta}. The
  topological numbers $(\nu_{0},\nu_{\pi})$ for the negative and
  positive $x$ regions are shown in the figures.  A letter ``N'' means
  that the topological number for the corresponding quasienergy cannot
  be defined because of the gap closing.  (b) The position dependence
  of the probability distribution $P(x,t)$ at $t=100$.  The
  eigenvalues of the timestep operators are shown in the insets as in
  Figs.~\ref{fig:Probability_Usingle} (b-1) and (b2).  }
\label{fig:Probability_Usplit_Hadamard}
\end{figure*}

We next consider the split-step Hadamard quantum walk, as defined in
Eq.~\eqref{eq:U_split}, with two coin operators
$C_\text{H}[\theta_{1,2}(x)]$ applied during one timestep.  We
emphasize that this quantum walk has been realized in an optical
experiment\cite{kitagawa_observation}.
We take a translation invariant bulk,
with $\theta_j(x)=\theta_j$, for both $j=1,2$,   
in a chiral timeframe, as defined by Eq.~\eqref{eq:U_split_chiral}. 
The operator $F_B$ reads 
\begin{align}
F_B &= e^{-i \chi} 
e^{-i\theta_1/2 \cdot \sigma_2}
\begin{pmatrix}
1&0\\0 & e^{i (k+2\chi)} 
\end{pmatrix}
e^{-i\theta_2/2 \cdot \sigma_2}.
\label{eq:F_B}
\end{align}
Again, the parameter $\chi$ does not affect the winding numbers, then we
can set $\chi=0$.

The calculation of the topological invariants follows the same lines
as for the simple Hadamard walk. We substitute Eq.~(\ref{eq:F_B}) into
Eqs.~\eqref{eq:F_components} and \eqref{eq:nu_phi_pi_def}. We obtain
\begin{subequations}
\begin{align}
\nu_{\pi} &= \frac{1}{2\pi i} \int_{-\pi}^{\pi} dk 
\frac{d}{dk} \text{ln } 
\left( s_{\theta_1+}s_{\theta_2+}\, - e^{ik} s_{\theta_1-}s_{\theta_2-}  \right);\\
\nu_{0} &= \frac{1}{2\pi i} \int_{-\pi}^{\pi} dk 
\frac{d}{dk} \text{ln } 
\left(\, s_{\theta_1-}s_{\theta_2+}\, + e^{ik} s_{\theta_1+}s_{\theta_2-}  \right),
\end{align}
\end{subequations}
using the shorthands defined in Eq.~\eqref{eq:s_pm_def}, and bearing
in mind that the quasienergies are displaced by $\pi$ instead of
$\pi/2$ as for the simple Hadamard walk.

Using the same logic as for the simple Hadamard walk for the winding
numbers gives us
\begin{subequations}
\label{eqs:nu_splitstep}
\begin{align}
\nu_{\pi} &= \left(\text{sign}  
[- \sin \theta_1 - \sin \theta_2] + 1 \right)/2; \\
\nu_{0} &= \left(\text{sign}  [ \sin \theta_1 - \sin \theta_2]
+ 1 \right)/2. 
\end{align}
\end{subequations}
The phase diagram in Fig.\ \ref{fig:phase-diagram} of the topological
numbers $\nu_0$ and $\nu_\pi$ at quasienergies $\varepsilon=0,\pi$,
respectively, of $U_B(\theta_1, \theta_2)$.

\subsubsection*{Numerical examples}

We illustrate the topological properties of the split-step Hadamard
walk using two parameter sets, 
\begin{align}
\label{eq:splitstep_example_gamma}
 {\bm \theta}_\gamma &: \{\theta_{1-}=\theta_{2-}=-\pi/4;\ 
 \theta_{1+}=\theta_{2+}=+\pi/4\};\\
\label{eq:splitstep_example_delta}
 {\bm \theta}_\delta &: \{\theta_{1-}=-\theta_{2-}=-\pi/4;\ 
 \theta_{1+}=\theta_{2+}=+\pi/4\},
\end{align}
where 
\begin{equation*}
\theta_{1(2)-} := \theta_{1(2)}(x\le -1),  \quad\theta_{1(2)+} :=
 \theta_{1(2)}(x \ge 0).
\label{eq:theta_- theta_+ 12}
\end{equation*}
Because only $\theta=\pm \pi/4$ is employed, the quantum walk
$U_B({\bm \theta}_{\gamma,\delta})$ would be called the split-step
Hadamard walk.  As indicated by the triangles in
Fig.\ \ref{fig:phase-diagram}, the parameters of the sets ${\bm
  \theta}_{\gamma,\delta}$ are located on the red solid and blue
dashed lines indicating the quasienergy gap closing around
$\varepsilon=0$ and $\pi$, respectively.  In case of ${\bm
  \theta}_\gamma$, the quasienergy gap around $\varepsilon=0$
vanishes, while the other gap around $\varepsilon=\pi$ is still open.
Since the topological numbers $\nu_\pi$ differ in the positive and
negative $x$ regions for $U_B({\bm \theta}_\gamma)$, the edge states
should exist.
However, in case of ${\bm \theta}_\delta$, 
parameters of the negative (positive) $x$ region locates on the
blue dashed (red solid) line. This results in no more energy gaps in the whole system. 
Then, we predict no edge states for $U_B({\bm \theta}_\delta)$. 

We confirm these predictions from the phase diagram by numerical simulations as shown in Fig.\
\ref{fig:Probability_Usplit_Hadamard}. In case of $U_B({\bm
\theta}_\gamma)$, we confirm the edge state at the quasienergy
$\varepsilon=\pi$ as well as the gap closing around $\varepsilon=0$. 
We also numerically confirm that no energy gaps emerge for $U_B({\bm
\theta}_\delta)$, and then no edge states.

\section{Interpretations of hidden topological invariants in experiment}
\label{sec:experiment}

Finally, we resolve the hidden topological invariant found in the
photonic quantum walk experiment in
Ref.\ \onlinecite{kitagawa_observation} where the time-evolution
operator is the one given in Eq.\ (\ref{eq:U_split}).  We consider two
parameter sets which are also investigated in the
experiment\cite{kitagawa_observation}:
\begin{align}
 {\bm \theta}_{e1} : \{&\theta_{1-}=0,\ \theta_{2-}=+\pi/4; \nonumber \\
&\quad \theta_{1+}=+7\pi/16,\ \theta_{2+}=+\pi/4\},
\label{eq:theta_exp1}\\
 {\bm \theta}_{e2} : \{&\theta_{1-}=-5\pi/16,\ \theta_{2-}=0; \nonumber \\
&\quad \theta_{1+}=+13\pi/16,\ \theta_{2+}=0\},
\label{eq:theta_exp2}
\end{align}
In Ref.\ \onlinecite{kitagawa_observation}, it is reported that 
the parameter spaces with the topological numbers $(1,0)$ and $(0,1)$
in Fig.\ \ref{fig:phase-diagram} have the topological number $0$ and 
the regions with the topological number  $(0,0)$ and $(1,1)$
in Fig.\ \ref{fig:phase-diagram} have the topological number $1$.
Thereby, in the case of ${\bm \theta}_{e1}$, 
the topological numbers of the positive and negative $x$ regions differ
by $1$, and then edge states are expected. 
In the case of ${\bm \theta}_{e2}$, 
the topological numbers in both regions are zero, and then edge states
are not expected.
However, edge states are observed in both cases in the experiment.

\begin{figure}[t]
\includegraphics[width=6.5cm]{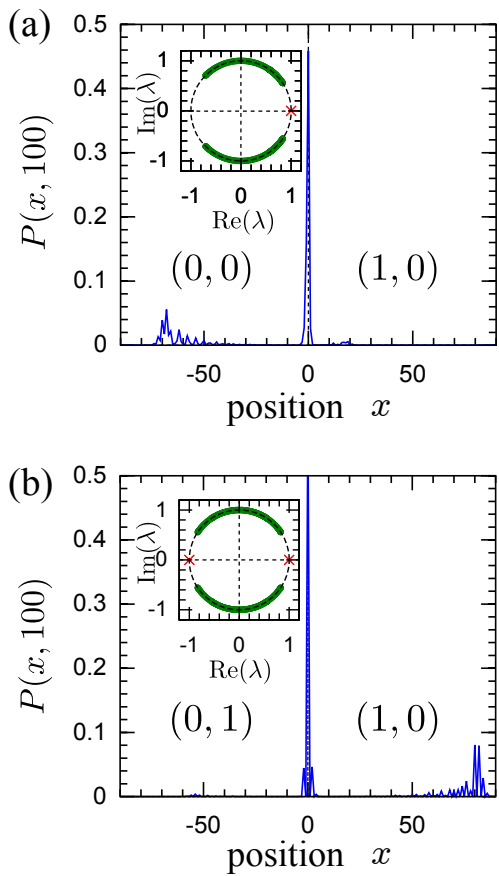}
\caption{(Color online)
The position dependence of the probability distribution $P(x,t)$ at
 $t=100$ is shown in (a) and (b) for the split-step walks
 $U_B({\bm \theta}_{e1})$ and $U_B({\bm \theta}_{e2})$, respectively.
The topological numbers $(\nu_{0},\nu_{\pi})$ for the negative
 and positive $x$ regions are shown in the figures.
The meaning of insets  is the same as those in
 Figs. \ref{fig:Probability_Usingle} (b-1) and (b2).
}
\label{fig:Probability_Usplit}
\end{figure}

In Fig.\ \ref{fig:Probability_Usplit} (a) and (b),
we show the probability distribution $P(x,t)$ 
at $t=100$ of the walks $U_B({\bm \theta}_{e1})$ and
$U_B({\bm \theta}_{e2})$, respectively, starting from  
the initial state
\begin{equation*}
 |\psi(t=0)\rangle := |+\rangle \otimes |0\rangle,
\end{equation*}
which also coincides with the experimental.
The corresponding eigenvalues of $U_B({\bm \theta}_{e1})$ and
$U_B({\bm \theta}_{e2})$ are shown in the insets. 
We confirm that the split-step walks $U_B({\bm \theta}_{e1})$ and
$U_B({\bm \theta}_{e2})$ exhibit edge states at $\varepsilon=0$  and
$\varepsilon=0,\pi$, respectively.

Now we look at the phase diagram in Fig.\ \ref{fig:phase-diagram}.
The phase diagram predicts that the quantum walk $U_B({\bm
  \theta}_{e1})$ should have an edge state at $\varepsilon=0$ because
the regions for $x\le -1$ and $x \ge 0$ have the topological numbers
$(\nu_0,\nu_\pi)=(0,0)$ and $(1,0)$, respectively. Furthermore,
in the case of $U_B({\bm \theta}_{e2})$, the edge states should appear at
quasienergies $\varepsilon=0,\pi$ because the regions for $x\le -1$
and $x \ge 0$ have the topological numbers $(0,1)$ and $(1,0)$,
respectively.  These theoretical results are completely consistent
with the numerical results in Fig.\ \ref{fig:Probability_Usplit} and
observations in the experiment\cite{kitagawa_observation}.  Thereby,
we have succeeded in explaining the hidden topological invariant found in the
experiment by the phase diagram Fig.\ {\ref{fig:phase-diagram}} which
is derived by establishing the relation between the rotation and
Hadamard matrices.

\section{Discussions and conclusion}
\label{sec:discussion}

Comparing Fig.\ \ref{fig:Probability_Usingle} (a-1) and (a-2) with
Fig.\ \ref{fig:Probability_Usplit_Hadamard} (a), we notice that the
the contour maps of the probability distribution $P(x,t)$ in the
former ones are sparser and checkerboard like.  The origin of this is
a sublattice symmetry of the time-evolution operator defined as
\begin{equation}
 {\Gamma_S} U {\Gamma_S} = -U,
\label{eq:sublattice}
\end{equation}
where 
\begin{equation*}
{\Gamma_S} :=  (\sum_{x\in \text{even}}|x\rangle \langle x|
- \sum_{x\in \text{odd}}|x \rangle \langle x|)\otimes \sigma_0.
 \end{equation*}
The single step walk $U_A$ retains sublattice symmetry, while the
split-step walk $U_B$ generally does not.
This symmetry constrains the walker at every timestep to hop from the
even sublattice ($x$ even) to the odd sublattice ($x$ odd), leading to
the checkerboard pattern of $P(x,t)$.

If a quantum walk has sublattice symmetry, as defined by
Eq.~\eqref{eq:sublattice}, every eigenstate at quasienergy
$\varepsilon$ must have a sublattice partner at quasienergy
$\varepsilon+\pi$\cite{obuse_delocalization,shikano2010localization}.
This explains why the single step Hadamard walk has edge states at
$\varepsilon=\pm \pi/2$ appearing simultaneously. We note that, due to
the specific value of $\theta_2=0$ in the parameter set ${\bm
  \theta}_{e2}$, the split-step walk $U_B({\bm \theta}_{e2})$ has
sublattice symmetry.  Therefore, the edge states appearing at
$\varepsilon=0,\pi$ in the inset of Fig.\ \ref{fig:Probability_Usplit}
(b) are sublattice symmetric partners of each other.

The results of this manuscript can be applied straightforwardly to
quantum walks where the generalized Hadamard coin is replaced by
operators $e^{-i\theta \sigma_2} \sigma_j$, with $j=1,2$, instead of
$j=3$.  In the case of $j=2$, we again obtain a chiral symmetric walk: in
analogy with Eq.\ (\ref{eq:def_chi_phi}), we have
\begin{eqnarray}
  e^{-i\theta \sigma_2} \sigma_2& = &
 e^{-i\phi} e^{-i(\theta+\eta)\sigma_2},
\label{eq:Cr_sigma2}
\end{eqnarray}
where $\eta=-\phi=\pi/2$.
Thereby, $\sigma_2$ shifts the angle of the rotation matrix and the
quasienergy by $\pi/2$, but does not affect chiral symmetry. The other
case, $j=1$, is easily understood because $\sigma_1= -i \sigma_2
\sigma_3$. Using Eqs.\ (\ref{eq:def_chi_phi}) and
(\ref{eq:Cr_sigma2}), we obtain
\begin{equation}
 e^{-i\theta \sigma_2}  \sigma_1 =  e^{-i\phi}  e^{-i(\theta+\eta)\sigma_2} 
\cdot e^{-i \chi \sigma_3},
\end{equation}
where $\eta=\chi=-\phi=\pi/2$ are global and fixed parameters. Thereby,
$\sigma_1$ also does not break chiral symmetry.

In Ref.~\onlinecite{kitagawa_observation}, two
topological invariants, $Q^0$ and $Q^\pi$ are defined that are
associated with a boundary between two bulks. These invariants are
nothing but the topologically protected number of 0 and $\pi$
quasienergy edge states localized at the boundary. The topological
invariants $\nu_0$ and $\nu_\pi$ we define in this paper are
associated with the bulks. By the arguments detailed in
Ref.~\onlinecite{asboth_2013}, the change in $\nu_0$ ($\nu_\pi$) as we
cross from the left bulk to the right predicts the value of $Q^0$
($Q^\pi$): this is the bulk--boundary correspondence for
one-dimensional Hadamard quantum walks.

The formulas for the topological invariants in this paper assumed
translation invariance in the bulk, so that the quasi-momentum $k$ is
a good quantum number. For disordered quantum walks, e.g., where
$\theta(x)$ is a random function of position, there are alternative
formulations of the topological invariants, based on the scattering
matrices\cite{scattering_walk2014, Rakovszky2015}. These can be
applied to disordered Hadamard walks using the mapping we presented in
this paper.

In summary, we have studied the topological phases of the
one-dimensional Hadamard quantum walk. We have generalized the
definition of chiral symmetry, and provided a sufficient requirement
for quantum walks to obey this symmetry, in
Eq.~\eqref{eq:FG}. Employing the generalized definition,
one-dimensional Hadamard quantum walks have chiral symmetry,
and the corresponding topological invariants, which characterize the
topological phases, can be calculated.  We have used this result to
reveal the topological invariants behind a recent photonic quantum
walk experiment\cite{kitagawa_observation}. Our results add to the
growing body of knowledge on the topological phases of quantum walks
and Floquet topological insulators.

\section{Acknowledgments}

We thank T.\ Endo and N.\ Konno for helpful discussions.  H.\ O.\ was
supported by Grant-in-Aid (No.\ 25800213 and N.\ 25390113) from the Japan
Society for Promotion of Science, J.~K.~A.~by the Hungarian Academy of
Sciences (Lend\"ulet Program, LP2011-016), by the Hungarian Scientific
Research Fund (OTKA) under Contract No. NN109651, and by the Janos
Bolyai scholarship.  The work of N.\ K.\ was supported by Grant-in-Aid
(No.\ 25400366 and No.\ 15H05855) from the Japan Society for Promotion of Science.

\appendix
\section{Alternative calculation of topological numbers}
\label{sec:appendix}
In Sec.\ \ref{sec:topological number}, we calculate the topological
numbers of the quantum walks based on the method developed in
Ref.\ \onlinecite{asboth2014chiral}.  Here, for completeness, we
present the calculation following the method of
Ref.\ \onlinecite{asboth_2013}.  The two methods are equivalent, we
present this derivation for pedagogical reasons.

\subsection{Single-step Hadamard walk}
First, we focus on the single-step Hadamard walk.  From
Eq.\ (\ref{eq:UA_single_chiral}), the chiral symmetric form of the
single-step Hadamard walk is written as
\begin{eqnarray}
U_A &=& e^{-i \phi} \cdot F_A \cdot \Gamma F_A^{-1} \Gamma,\\
F_A &=& 
e^{-i\theta(x)/2 \cdot \sigma_2} \cdot
e^{-i\chi/2 \cdot \sigma_3 } 
S_-,\nonumber \\
\Gamma F_A^{-1} \Gamma &=&
S_+ \cdot
e^{-i\chi/2 \cdot \sigma_3 }  \cdot
e^{-i\theta(x)/2 \cdot \sigma_2 }
,\nonumber
\end{eqnarray}
where $\chi=-\phi=\pi/2$.
As explained in Ref.\ \onlinecite{asboth_2013}, if a time-evolution
operator of the quantum walk $U$ has chiral symmetry, 
another chiral symmetric time-evolution operator $U^\prime$ can be
identified only at the different ``timeframe''. 
The one for $U_A$, thus $U_A^\prime$, is given by
\begin{eqnarray}
U_A^{\prime} &=& e^{-i \phi}\cdot \Gamma F_A^{-1} \Gamma \cdot F_A.
\end{eqnarray}
Since the phase $\phi$ only shifts the quasienergy, we proceed in our calculations
by setting $\phi=0$ on the above equations during the
calculation and shift the quasienergy by $\pi/2$ at the end. 
In the momentum representation (by assuming the constant $\theta$),
we have
\begin{eqnarray*}
U_A(k)  &=& 
e^{-i \theta/2 \cdot \sigma_2} 
\begin{pmatrix}
e^{-i (k+\chi)} & 0 \\
0 & e^{ik(k+\chi)}
\end{pmatrix}
e^{-i \theta/2 \cdot \sigma_2} \\
&=&
\cos(k+\chi) c_\theta \sigma_0 - i c_k s_\theta \sigma_2 - i \sin(k+\chi) \sigma_3,
\end{eqnarray*}
and 
\begin{eqnarray*}
U_A^\prime(k)  &=& 
\begin{pmatrix}
e^{-i (k+\chi)} & 0 \\
0 & 1
\end{pmatrix}
e^{-i \theta \cdot \sigma_2} 
\begin{pmatrix}
1 & 0 \\
0 & e^{i(k+\chi)}
\end{pmatrix}
\\
&=&
\cos(k+\chi) c_\theta \sigma_0 - i s_\theta \sigma_2 - i \sin(k+\chi) c_\theta \sigma_3.
\end{eqnarray*}
Here, we use the shorthands
\begin{equation*}
 c_\theta \equiv \cos(\theta), \quad s_\theta \equiv \sin(\theta).
\end{equation*}
The factor $\chi$ only shifts
the momentum $k$, which does not change the topological number. Then, we
set $\chi=0$ in the following. 

Applying a unitary transform so that the time evolution operators 
have chiral symmetry in the basis that the chiral symmetry operator is
diagonal, i.e., $\Gamma=\sigma_3$, we have
\begin{eqnarray*}
\tilde{U}_A(k)  &=& e^{i \pi/4 \cdot \sigma_2} U_A(k) e^{-i \pi/4 \cdot \sigma_2}\\
&=&
c_k c_\theta \sigma_0 + i s_k \sigma_1 - i c_k s_\theta \sigma_2 ,
\end{eqnarray*}
and 
\begin{eqnarray*}
\tilde{U}_A^\prime(k)  &=& e^{i \pi/4 \cdot \sigma_2} U_A^\prime(k) e^{-i \pi/4 \cdot \sigma_2}\\
&=&
c_k c_\theta \sigma_0 + i s_k c_\theta \sigma_1 -i s_\theta \sigma_2.
\end{eqnarray*}
Since the coefficients of the $\sigma_0$ term of $\tilde{U}_A(k)$ and
$\tilde{U}_A^\prime(k)$ are the same, both operators have a
common eigenvalue
\begin{equation*}
 \lambda_{A,\pm} = e^{\pm i \omega_A},\quad \sin(\omega_A) = \sqrt{1-(c_k
  c_\theta)^2} \ge 0.
\end{equation*}
The corresponding eigenvectors $\ket{\psi_\pm}$ and
$\ket{\psi_\pm^\prime}$ of  $\tilde{U}_A(k)$ and
$\tilde{U}_A^\prime(k)$, respectively, also have the similar structures
\begin{equation}
 \ket{\psi_{A,\pm}} = \frac{1}{\sqrt{2}} 
\begin{pmatrix}
\mp i e^{i \varphi_A(k)} \\ 1
\end{pmatrix},
 \ket{\psi^\prime_{A,\pm}} = \frac{1}{\sqrt{2}} 
\begin{pmatrix}
\mp i e^{i \varphi_A^\prime(k)} \\ 1
\end{pmatrix},
\label{eq:eigenvector}
\end{equation}
but with different phase factors
\begin{subequations}
\begin{align}
 e^{i\varphi_A(k)} &= (-c_k s_\theta + i s_k)/\sin(\omega_A),\\
 e^{i\varphi_A^\prime(k)} &= (-s_\theta + i s_k
  c_\theta)/\sin(\omega_A).
\end{align}
\label{eq:phi_A}
\end{subequations}
The winding number is defined through the Berry phase,
\begin{eqnarray}
\nu \equiv \frac{1}{i \pi} \int dk \bra{\psi} d/dk \ket{\psi}.
\label{eq:winding}
\end{eqnarray}
Substituting eigenvectors in Eq.\ (\ref{eq:eigenvector}), the
winding numbers $\nu$ and $\nu^\prime$ of $U_A$ and $U_A^\prime$,
respectively, become
\begin{equation*}
 \nu = \frac{1}{2\pi}\oint d\varphi_A(k), \quad  \nu^\prime = \frac{1}{2\pi}\oint d\varphi_A^\prime(k).
\end{equation*}
Thereby, the winding numbers are determined from the trace of
 $\varphi_A(k)$ and $\varphi_A^\prime(k)$ as $k$ is changed from $0$ to $2\pi$.
Considering Eq.\ (\ref{eq:phi_A}), this  gives the following results,
\begin{eqnarray*}
\nu= \left\{
\begin{array}{rr}
-1 & (0<\theta<\pi)\\
1 & (-\pi<\theta<0)
\end{array}
\right.,
\end{eqnarray*}
and 
\begin{equation*}
\nu^\prime = 0. 
\end{equation*}
Finally by using a formula derived in Ref.\ \onlinecite{asboth_2013} in
order to calculate the topological numbers for
quasienergies $\varepsilon=\phi$  and $\pi+\phi$
\begin{equation}
 \nu_{\phi} = \frac{\nu^\prime + \nu}{2}, \quad \nu_{\pi+\phi} = \frac{\nu^\prime-\nu}{2},
\label{eq:nu_0 nu_pi}
\end{equation}
we obtain 
\begin{equation}
 (\nu_{-\pi/2}, \nu_{+\pi/2}) =
\left\{
\begin{array}{rl}
(-1/2, +1/2) & \quad \text{for } 0<\theta<\pi\\
(+1/2,-1/2) & \quad \text{for }-\pi<\theta<0.
\end{array}
\right.
\label{eq:result U_A}
\end{equation}
Since the global shift of topological numbers does not alter the
argument of the bulk-edge correspondence, we confirm the consistent
result with Eq.\ (\ref{eq:nu for the single step Hadamard}) in Sec.\
\ref{sec:topological number} by shifting numbers in the right side of Eq.\ (\ref{eq:result U_A}) by $1/2$.

\subsection{Split-step Hadamard walk}

In the case of the split-step Hadamard walk, we have the following two
chiral symmetric time-evolution operators
\begin{eqnarray}
U_B &=  e^{-i 2\phi} \cdot F_B \cdot \Gamma F_B^{-1} \Gamma,\\
U_B^\prime &=  e^{-i 2\phi} \cdot \Gamma F_B^{-1} \Gamma \cdot F_B,
\end{eqnarray}
where
\begin{eqnarray*}
F_B &= 
e^{-i\theta_1(x)/2 \cdot \sigma_2 } 
S_- \cdot
e^{-i\chi \sigma_3} \cdot
e^{-i\theta_2(x)/2 \cdot \sigma_2 },
\\
\Gamma F_B^{-1} \Gamma &= 
e^{-i\theta_2(x)/2 \cdot \sigma_2 } \cdot
e^{-i\chi \sigma_3} 
S_+ \cdot
e^{-i\theta_1(x)/2 \cdot \sigma_2 },
\end{eqnarray*}
with $\chi=-\phi=\pi/2$. Again we set $\phi=0$ of $U_B$ and
$U_B^\prime$ and shift the quasienergy by $\pi$ at the end of the
calculation. 
We derive the time evolution operators in the  momentum space
representation as
\begin{eqnarray}
 U_B(k) &=&
  [\cos(k+2\chi)c_{\theta_2}c_{\theta_1}-s_{\theta_2}s_{\theta_1}]\sigma_0
  \nonumber \\
&-&i[\cos(k+2\chi) c_{\theta_2}s_{\theta_1} +
 s_{\theta_2}c_{\theta_1}]\sigma_2
\nonumber\\
&-&i\sin(k+2\chi) c_{\theta_2} \sigma_3,
\label{eq:U_B}
 \end{eqnarray}
and 
\begin{eqnarray*}
 U_B^\prime(k) &=&
  [\cos(k+2\chi)c_{\theta_2}c_{\theta_1}-s_{\theta_2}s_{\theta_1}]\sigma_0
  \nonumber \\
&-&i[\cos(k+2\chi) c_{\theta_1}s_{\theta_2} +
 s_{\theta_1}c_{\theta_2}]\sigma_2
\nonumber\\
&-&i\sin(k+2\chi) c_{\theta_1} \sigma_3.
\label{eq:U_B'}
 \end{eqnarray*}
Comparing the above two equations, we notice that
$U_B(k)$ and $U_B^\prime(k)$ are identical only by switching $\theta_1$ and $\theta_2$. 
This means that results for $U_B(k)$ are immediately applied to those for $U_B^\prime(k)$ by switching $\theta_1$ and
$\theta_2$.
Thereby, we present calculations only for $U_B(k)$ hereafter.

Similarly to the single-step Hadamard walk case, we can set $\chi=0$ in
Eq.\ (\ref{eq:U_B}) and
apply the unitary transformation, we have
\begin{eqnarray*}
 \tilde{U}_B(k) &=&  
e^{i \pi/4 \cdot \sigma_2} U_B(k) e^{-i \pi/4\cdot \sigma_2} \nonumber \\
&=&  [\cos(k)c_{\theta_2}c_{\theta_1}-s_{\theta_2}s_{\theta_1}]\sigma_0
  \nonumber \\
&+&i\sin(k) c_{\theta_2} \sigma_1
\nonumber \\
&-&i[\cos(k) c_{\theta_2}s_{\theta_1} +
 s_{\theta_2}c_{\theta_1}]\sigma_2.
 \end{eqnarray*}
The eigenvalue of $\tilde{U}_B(k)$ is
\begin{eqnarray*}
 \lambda_{B,\pm} &=& e^{\pm i \omega_{B}},\\
 \sin(\omega_B) &=&
  \sqrt{1-[\cos(k)c_{\theta_2}c_{\theta_1}-s_{\theta_2}s_{\theta_1}]^2}
  \ge 0,
\end{eqnarray*}
and the corresponding eigenvector is
\begin{eqnarray}
 \ket{\psi_{B,\pm}} &=& \frac{1}{\sqrt{2}} 
\begin{pmatrix}
\mp i e^{i \varphi_B(k)} \\ 1
\end{pmatrix},
\label{eq:eigenvector U_B}\\
 e^{i\varphi_B(k)} &=& 
\frac{-[\cos(k) c_{\theta_2} s_{\theta_1}+s_{\theta_2} c_{\theta_1}] 
+ i\sin(k) c_{\theta_2}}
{\sin(\omega_B)}.
\nonumber
\end{eqnarray}

Substituting Eq.\ ({\ref{eq:eigenvector U_B}}) into Eq.\ (\ref{eq:winding}), the winding number $\nu$ of $U_B$
is summarized as follows:
when 
\begin{equation}
 \sin^2(\theta_1) - \sin^2(\theta_2) > 0,
\end{equation}
\begin{eqnarray}
\nu= \left\{
\begin{array}{rr}
-1 & (0<\theta_1<\pi)\\
1 & (-\pi<\theta_1<0)
\end{array}
\right.,
\end{eqnarray}
otherwise
\begin{equation}
 \nu=0.
\end{equation}
As we mentioned, the winding number $\nu^\prime$ of $U_B^\prime(k)$ is
given by switching $\theta_1$ and $\theta_2$ in the above results for $\nu$.

Finally, we obtain the consistent phase diagram with that of Fig.\
\ref{fig:phase-diagram} in Sec.\ \ref{sec:topological number}, 
by substituting  $\nu$ and $\nu^\prime$ of $U_B$ and $U_B^\prime$, respectively, into 
Eq.\ (\ref{eq:nu_0 nu_pi}), taking account of the quasienergy shift by
$\pi$, and $1/2$ shift of the topological numbers.

\bibliography{walkbib}{}

\end{document}